\documentclass[10pt,twocolumn,letterpaper]{article}

\usepackage[pagenumbers]{cvpr} 

\usepackage[dvipsnames]{xcolor}

\definecolor{cvprblue}{rgb}{0.21,0.49,0.74}
\usepackage[pagebackref,breaklinks,colorlinks,citecolor=cvprblue]{hyperref}

\def\paperID{} 
\def\confName{CVPR}
\def\confYear{***}

\title{Interactive3D: Create What You Want by Interactive 3D Generation}

\author{%
  Shaocong Dong$^{1}$\footnotemark[1], ~~Lihe Ding$^{2,4}$\footnotemark[1], ~~Zhanpeng Huang$^{3}$, ~~Zibin Wang$^{3}$, \\
  ~~Tianfan Xue$^{2}$\footnotemark[2], ~~Dan Xu$^{1}$\footnotemark[2] \\
  {$^1$}Hong Kong University of Science and Technology 
  ~~{$^2$}The Chinese University of Hong Kong \\
  {$^3$}SenseTime Research ~~{$^4$}Shanghai AI Laboratory \\
  {\tt\small\{sdongae, danxu\}@cse.ust.hk, \{dl023, tfxue\}@ie.cuhk.edu.hk}\\
  {\tt\small \{wangzb02, yiyuanzhang.ai\}@gmail.com, \{huangzhanpeng\}@sensetime.com}\\
}

\usepackage{cuted}
\usepackage{capt-of}
\usepackage{wrapfig}
\usepackage[accsupp]{axessibility} 

\begin{document}
\maketitle
\begin{strip}\centering
\vspace{-16mm}
\includegraphics[width=\textwidth]{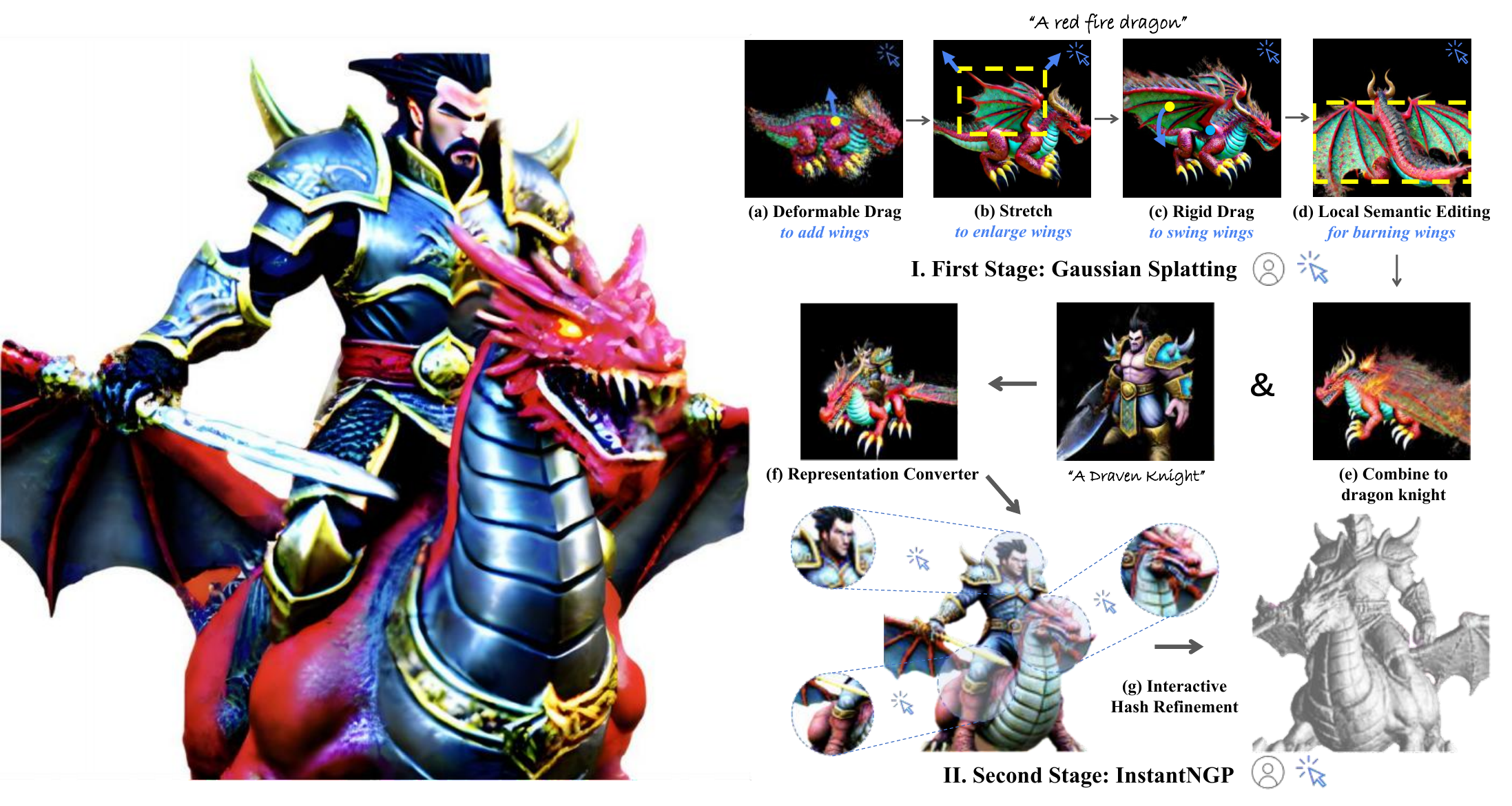}
\vspace{-7mm}
\captionof{figure}{\textbf{Interactive 3D Generation}.~\textbf{The first stage} involves using Gaussian ellipsoids for creating a base model where users can interact through different operations, such as deformable dragging to add features like wings, stretching to enlarge parts, rigid dragging to reposition elements, and local semantic editing to apply specific visual effects, e.g., making wings appear aflame. \textbf{The second stage} shows the conversion of this proposed Gaussian representation into an InstantNGP structure followed by an Interactive Hash Refinement process, allowing for further detailed enhancements. It demonstrates the framework's capability to merge and refine complex generations.
\vspace{-3.5mm}
\label{fig:teaser}}
\end{strip}

\renewcommand{\thefootnote}{\fnsymbol{footnote}}
\footnotetext[1]{Equal contribution.}
\footnotetext[2]{Corresponding author.}

\begin{abstract}
\vspace{-1mm}
3D object generation has undergone significant advancements, yielding high-quality results. However, fall short of achieving precise user control, often yielding results that do not align with user expectations, thus limiting their applicability. User-envisioning 3D object generation faces significant challenges in realizing its concepts using current generative models due to limited interaction capabilities. Existing methods mainly offer two approaches: (i) interpreting textual instructions with constrained controllability, or (ii) reconstructing 3D objects from 2D images. Both of them limit customization to the confines of the 2D reference and potentially introduce undesirable artifacts during the 3D lifting process, restricting the scope for direct and versatile 3D modifications.~In this work, we introduce \textbf{Interactive3D}, an innovative framework for interactive 3D generation that grants users precise control over the generative process through extensive 3D interaction capabilities. Interactive3D is constructed in two cascading stages, utilizing distinct 3D representations. The first stage employs Gaussian Splatting for direct user interaction, allowing modifications and guidance of the generative direction at any intermediate step through \textbf{(i)} Adding and Removing components, \textbf{(ii)} Deformable and Rigid Dragging, \textbf{(iii)} Geometric Transformations, and \textbf{(iv)} Semantic Editing. Subsequently, the Gaussian splats are transformed into InstantNGP. We introduce a novel \textbf{(v)} Interactive Hash Refinement module to further add details and extract the geometry in the second stage. Our experiments demonstrate that proposed Interactive3D markedly improves the controllability and quality of 3D generation. Our project webpage is available at \url{https://interactive-3d.github.io/}.
\end{abstract}

\vspace{-8pt}
\section{Introduction}
\label{sec:intro}

Recent advancements~\cite{metzer2022latent,rombach2022high,saharia2022photorealistic} in 2D image generation, exemplified by approaches, such as diffusion models trained on extensive text-image paired datasets (e.g., LAION-series~\cite{schuhmann2021laion}), have made significant strides in aligning generated images with textual prompts. Despite this success, achieving precise control over image generation to meet complex user expectations remains a severe challenge. ControlNet~\cite{zhang2023controlnet} addresses this by modifying foundational 2D diffusion models with fine-tuning on specific conditional datasets, offering a subtle control mechanism guided by user-specific inputs.

On the other hand, 3D object generation, despite its promising  progress~\cite{poole2022dreamfusion, wang2023prolificdreamer}, confronts more intricate challenges than those encountered in 2D image generation. 
Although advancements have been observed from perspectives, including
training 3D diffusion models on direct 3D datasets~\cite{jun2023shap, nichol2022point}, and lifting 2D diffusion priors to 3D representations (e.g., NeRF~\cite{mildenhall2021nerf}) via techniques like SDS loss optimization~\cite{poole2022dreamfusion}, \textit{precise control over the generated objects has not been fully achieved}. The reliance on initial text prompts or 2D reference images severely limits the generation controllability and often results in lower quality. The text prompts lack specificity to convey complex 3D designs accurately; while the 2D reference images can inform 3D reconstruction, they do not capture the full depth of 3D structures, potentially leading to various unexpected artifacts. Moreover, personalization based on 2D images lacks the flexibility that can be offered by direct 3D manipulation.

A straightforward idea to achieve controllable 3D generation is to adapt the ControlNet to 3D generation. However, this strategy encounters significant obstacles: (i) the control signals for 3D are inherently more complex, making the collection of a conditioned 3D dataset exceptionally challenging when compared to the 2D paradigm; (ii) the absence of powerful foundational models in the 3D domain, like stable diffusion for 2D~\cite{metzer2022latent}, impedes the possibility of developing fine-tuning techniques at this time. These hurdles suggest the need for a different strategy. As a result, we are inclined to explore a novel question: \textit{can we directly integrate flexible human instructions into the 3D generation?}

\begin{figure*}[htbp]
  \setlength{\abovecaptionskip}{0cm}
  \centering
    \includegraphics[width=0.97\linewidth]{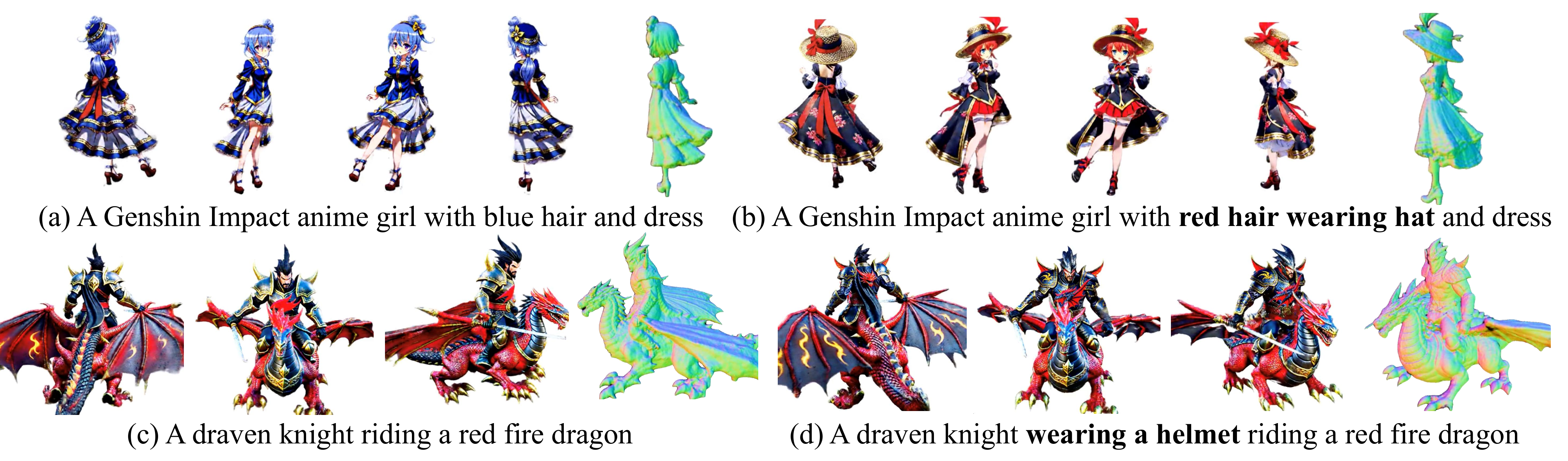}
  \caption{Qualitative generation results of the proposed Interactive3D. We achieve high-quality and controllable 3D generation.}
  \vspace{-15pt}
  \label{fig:refine}
\end{figure*}

To address the above-identified challenges, we introduce~\textit{Interactive3D}, a framework devised to facilitate user interaction with intermediate outputs of the generative process, enabling precise control over the generation and effective enhancement of generated 3D object quality. Our approach is characterized by a two-stage process leveraging distinct 3D representations. 
Specifically, the first stage utilizes the SDS loss~\cite{poole2022dreamfusion} to optimize a 3D object by Gaussian Splatting~\cite{kerbl20233d} representation, which allows for independent user modifications during the optimization process. The second stage transforms the Gaussian representation into InstantNGP~\cite{mueller2022instant} structures and applies proposed Interactive Hash Refinement for detailed textures and 3D geometry.

For the first stage, the Gaussian blobs enable direct user feedback, such as the addition and the removal of object components by manipulating the Gaussian blobs, as illustrated in the knight and dragon composition in~\cref{fig:teaser}, and a dragging operation based on the principles of DragGAN~\cite{tewari2023drag}. Users can select a source and a target point within the 3D space, which guide the Gaussian blobs (conceptualized as point clouds) from an original to a desired location using a \textit{motion-supervision loss}. Notably, our framework facilitates both deformable and rigid dragging of object elements through a \textit{rigid constraint loss}, allowing users to create new or adjust existing 3D model components.

Furthermore, our \textit{Interactive3D} allows for precise selection of subsets of Gaussian blobs, to which users can apply various transformations, e.g., the stretch operation shown in~\cref{fig:teaser} (b). By restricting the gradient flow to these subsets, we enable focused optimization of chosen 3D parts using modified text prompts, without altering other areas of the model. This strategy is termed as local semantic editing.
Its capability is exemplified in~\cref{fig:teaser} (d), where the wings are edited to simulate the effect of being ablaze. 
To enhance the guidance from user interactions during generation, we integrate an \textit{interactive SDS loss} with an adaptive camera zoom-in technique, significantly improving the optimization efficiency of the modified 3D areas.

At the beginning of the second stage, the modeled Gaussian representations are transformed into InstantNGP structures through a swift NeRF distillation technique. This strategic transformation synergies the strengths of both representations: Gaussian blobs, while more friendly for direct editing, face challenges in reconstructing high-quality 3D geometry; on the other hand, InstantNGP structures excel in providing a foundation for further geometry refinement and mesh extraction, as depicted in~\cref{fig:teaser} (f).
Given that the converted InstantNGP employs hash tables to associate learnable features with 3D grids, we have developed an innovative \textit{Hash Refinement Module}. This module enables the interactive enhancement and detailing of chosen areas within the base radiance fields, as demonstrated in~\cref{fig:teaser} (g). Specifically, we begin by extracting coarse occupancy grids from the initial radiance field. When a user selects an area for optimization, a candidate grid set is determined by the intersection of this area with the occupancy grids. These grids are then categorized into multi-level sets with finer resolutions than the original field, and multiple refinement hash tables are constructed to map the selected area to learnable refinement features, which are then fed into lightweight MLPs to capture residual colors and densities, concentrating detail enhancement on local surface regions.

Through our Interactive Hash Refinement module, users can precisely control the refinement process by selecting specific areas for optimization and adjusting the levels of representation details, such as the hash table levels and the capacities. Hence, users can sculpt their envisioned objects during generation by employing the array of interactive options provided by \textit{Interactive3D}.

In summary, our contributions are threefold: 
\begin{itemize}
    \item \textbf{Framework Deisgn.} We introduce a novel interactive 3D generation framework, \textit{Interactive3D}, which empowers users to precisely control the 3D generation process via direct instructions.

    \item \textbf{3D Representations.} We reveal that incorporating the Gaussian Splatting for user interaction and the InstantNGP for further geometry refinement and mesh extraction leads to higher-quality 3D generation. 

    \item \textbf{Generation Results.} With precise control and effective user feedback in the generation process, our method significantly improves the generation quality.

\end{itemize}



\section{Related Work}
\label{sec:formatting}
\subsection{3D Generative Models}
\vspace{-1mm}
3D Generation is a foundation task and has been studied widely. Early works focus on the 3D representations including voxels~\cite{wu2016learning, gadelha20173d, smith2017improved, henzler2019escaping, lunz2020inverse}, point clouds~\cite{achlioptas2018learning, mo2019structurenet, yang2019pointflow}, meshes~\cite{zhang2020image, shen2021deep, gao2022get3d} and implicit fields~\cite{chen2019learning, mescheder2019occupancy}. 
Recently, diffusion models~\cite{ho2020ddpm} have achieved great success in 2D content creation~\cite{nichol2021glide, saharia2022photorealistic, rombach2022high, ramesh2022hierarchical}, and there has been substantial research into 3D diffusion, which has significantly improved 3D generation. These 3D diffusion models can be divided into two directions: optimization-based~\cite{poole2022dreamfusion} and feed-forward methods~\cite{jun2023shap}. For the optimization-based methods, the DreamFusion series~\cite{jain2022zero, wang2022score, poole2022dreamfusion, lin2023magic3d, metzer2022latent} design an SDS loss based on probability density distillation,  which enables the use of a 2D diffusion model as a prior for optimization of a parametric image generator (e.g., rendered from NeRF~\cite{mildenhall2021nerf}). They optimize a 3D consistent radiance field via gradient descent such that its 2D renderings from random angles achieve low losses. These optimization-based methods can realize zero-shot and high-quality generation with several hours of optimization. Further works, such as Prolificdreamer~\cite{wang2023prolificdreamer} and Fantasia3d~\cite{chen2023fantasia3d}, achieve higher quality in 3D content generation by modifying the SDS loss. 
For feed-forward methods, PointE~\cite{nichol2022point} generates a single synthetic view using a text-to-image diffusion model, and then produces a 3D point cloud using a second diffusion model that is conditioned on the generated image. The follow-up work Shap-E~\cite{jun2023shap} trains a latent diffusion model on NeRF's parameters.
One-2-3-45~\cite{liu2024one} uses Zero-1-to-3~\cite{liu2023zero} to generate multi-view images, which are fed into a SparseNeuS~\cite{long2022sparseneus} for 3D generation. 
However, current methods~\cite{jain2022zero, wang2022score, poole2022dreamfusion, lin2023magic3d, metzer2022latent, chen2023fantasia3d, wang2023prolificdreamer} only rely on initial text prompts or reference images to 3D generation, restricting their controllability. In contrast, our framework introduces user interactions into the optimization, achieving flexible and controllable 3D generation.

\subsection{Gaussian Splatting}
The recent 3D representation Gaussian Splatting~\cite{kerbl20233d} has revolutionized the 3D rendering field by using a set of Gaussian blobs to represent the 3D scenes and splat them onto the image plane to obtain renderings. Compared with Neural Radiance Field (NeRF), Gaussian splatting models the 3D world by explicit Gaussian blobs, which are flexible and naturally suitable for human interactions and editing.

\subsection{InstantNGP}
To improve the quality and efficiency of NeRF,~\cite{mueller2022instant} proposes InstantNGP which utilizes multi-level hash tables to map between learnable features and 3D query positions. InstantNGP does not handle the hash conflict explicitly and makes the gradients to guide the optimization of features. In this case, the surface positions obtain larger gradients during training and dominate the feature updates. Thanks to the adaptive feature update by hash mapping, InstantNGP can represent the 3D world by more fine-grained grids and easily extract 3D geometries (e.g., mesh) from the continuous radiance field. However, InstantNGP is still an implicit representation and is hard to interact with or edit.
\section{The Proposed Interactive3D Approach}

\begin{figure*}[htbp]
  \setlength{\abovecaptionskip}{0cm}
  \centering
  \includegraphics[width=1.0\linewidth]{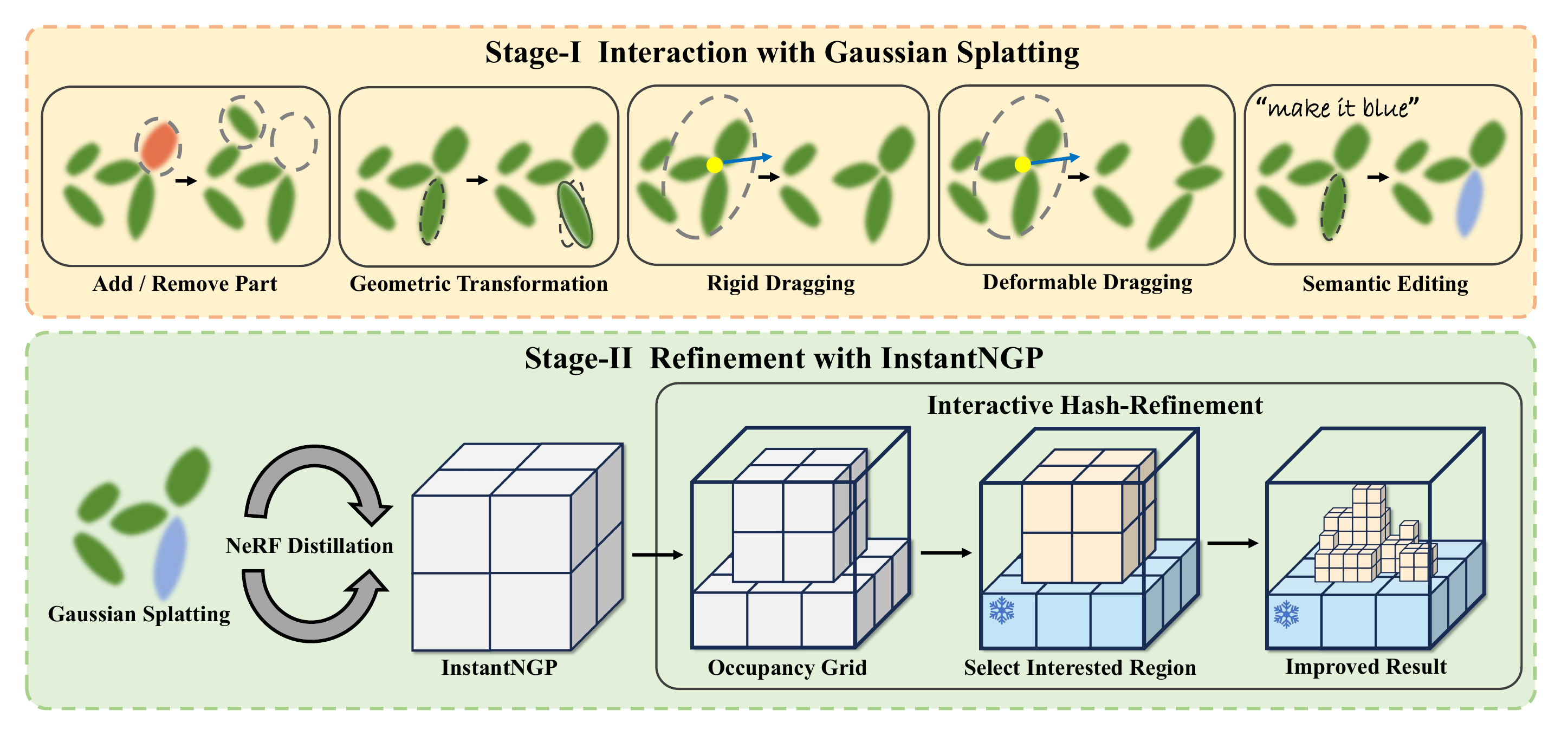}
  \vspace{-5pt}
  \caption{\textbf{The overall architecture of Interactive3D.} It contains two stages with distinct 3D representations: \textbf{(I)} Gaussian Splatting for flexible user interactions such as add/remove parts, geometry transformation, deformable or rigid dragging and semantic editing; \textbf{(II)} the Gaussian blobs are converted to InstantNGP using NeRF distillation and fine-tuned by our Interactive Hash Refinement Module.}
  \vspace{-9pt}
  \label{fig:overview}
\end{figure*}

Achieving Interactive 3D Generation is non-trivial and poses two core challenges: (i) we need to design an efficient and effective interactive mechanism for the 3D representation to control the generation process, and (ii) we need to obtain high-quality 3D outputs (e.g., the mesh) aligned with users' expectations.
As for the 3D representation, 
NeRF~\cite{mildenhall2021nerf} can output high-quality 3D objects, however, its implicit representation makes it hard to incorporate interactions. While the recent 3D Gaussian Splatting is naturally suitable for interaction due to the independence and explicitly of Gaussian blobs, it cannot output the commonly used mesh with high quality so far. Observing the complementarity of these two representations, we design a two-stage interactive 3D generation framework, where Stage I (\cref{sec:stage1}) adopts Gaussian Splatting to achieve flexible interactions and Stage II (\cref{sec:stage2}) converts the Gaussian blobs to InstantNGP for further geometry refinement and mesh extraction, as shown in~\cref{fig:overview}. Below we first introduce the first interactive generation stage with Gaussian Splatting.

\subsection{Interaction with Gaussian Splatting}
\label{sec:stage1}
In the first stage, we represent a 3D object as a set of $N$ Gaussian blobs $\mathcal{E} = \left \{(c_i, o_i, \mu_i, \Sigma_i) \right \}_{i}^{N}$, where $c_i, o_i, \mu_i, \Sigma_i$ represents the color, opacity, position, and the covariance of the $i$-th Gaussian, respectively. To incorporate interactions, we treat each Gaussian blob as a point and formulate a point cloud set $\mathcal{S} = \left \{\mu_i \right \}_{i}^{N}$. Thanks to the flexibility of treating Gaussian blobs as points, we can introduce various user interactions by explicitly modifying the point set $\mathcal{S}$, as discussed in \cref{sec:add}, \cref{sec:drag}, and \cref{sec:semantic}. Once the adjustment is finished, we adopt an interactive SDS loss to optimize the modified parts efficiently in \cref{sec:loss}.

\subsubsection{Adding and Removing Parts}
\label{sec:add}
To add parts, suppose we have two Gaussian blob sets $\mathcal{E}_1$ and $\mathcal{E}_2$, we can simply obtain the combined set by concatenating them. In practice, this interaction often happens when the user desires to combine two objects (e.g., knight and dragon as shown in~\cref{fig:teaser}).

To remove parts $P$, we can directly delete the Gaussian blobs within the part by:
\begin{equation}
    \mathcal{E}' = \mathcal{E} - \left \{ e_i \right \}_{i \in I(P)},
\end{equation}
where $I(p) = \left \{ i: \mu_i \in P \right \}$ indicates the indexes of Gaussian blobs in $P$.
Importantly, the determination of whether blob $e_i$ belongs to $P$ depends on how we define the parts. In practice, we offer two ways to select a part. The first one is defined in 2D images. Specifically, the users can choose a point on the rendered 2D images from Gaussian Splatting and utilize an off-the-shell segmentor (e.g., SAM~\cite{kirillov2023segment}) to obtain multi-view part masks. Subsequently, we project all the points in $\mathcal{S}$ onto all masked views and obtain the blobs with all projections within masks as the part blobs. We use 2 views by default.
However, such a 2D mask-based part selection needs to infer with a large-scale pretrained segmentation network, which may be accurate but not efficient. Furthermore, if we do not need a precise part set or we select parts at the early stage of generation (the rendered images may not be informative enough for SAM), we can directly select points in 3D. 

\subsubsection{Geometry Transformation}
\label{sec:geo}
Given that any interested part can be selected as discussed in ~\cref{sec:add}, we can take a further step to construct a bounding box $B$ for each $P$. In this way, all the traditional geometry transformations $\mathcal{T}$ such as rotation, translation, and stretching can be first applied to the part $P$ and then it is concatenated with the unchanged set as follows:
\begin{equation}
    \mathcal{E}_{\text{trans}} = \mathcal{T}(B(P)) + \mathcal{E}'.
\end{equation}

\subsubsection{Deformable and Rigid Dragging}
\label{sec:drag}
Although we have achieved the part-level geometry transformations, the users may prefer more flexible and direct interactions. Inspired by DragGAN~\cite{tewari2023drag}, we propose deformable and rigid 3D Dragging operations. The deformable dragging regards the local structure as a plasticine and aims to deform the geometry smoothly, e.g., dragging new wings from the dragon's back, while the rigid dragging treats the local region as a rigid part and forces the local structure unchanged. Specifically, we first select a source point $p_s$, a target point $p_t$, and a local region radius $r$. Then, the activated local object part can be written as:
\begin{equation}
\label{eq:local}
    P = \left \{ e_i | \| \mu_i - p_s \|_2 \leq r \right \}.
\end{equation}
Subsequently, we move all points with a minor offset along the direction from $p_s$ to $p_t$ before each optimization step:
\begin{equation}
    \mu_i' = \mu_i + \alpha\frac{(p_t - p_s)}{\| p_t - p_s \|_2},
\end{equation}
where $\alpha$~is~a~predefined hyper-parameter to control the movement step.~Then the modified blobs $\left \{ e_i' = (c_i, o_i, \mu_i', \Sigma_i) \right \}_{i=1}^{N}$ can render images through Gaussian Splatting. From the rendered images, we can compute the SDS loss $L_{\text{SDS}}$ following~\cite{poole2022dreamfusion, chen2023text}. Meanwhile, to encourage the part $P$ to move towards $p_t$ during optimization, we add a motion supervision loss following~\cite{tewari2023drag}:
\begin{equation}
    L_{\text{motion}} = \sum_{i \in I(P)}(\|\mu_i' - p_t\|_1),
\end{equation}
where $I(P)$ indicates a set of indexes of Gaussian blobs in $P$. Then we can compute the gradient and update the parameters of each Gaussian blob. Note that we do not need a point tracking process as in~\cite{tewari2023drag} since the 3D points are naturally tracked. Importantly, we apply densification and pruning operations as in~\cite{kerbl20233d} to fill the gaps and eliminate noises after the dragging operation. 

To achieve deformable or rigid dragging, we made two adjustments to the above process. First, for the deformable dragging, we continuously update the activated part $P$ by updating the current position of source point $p_s$ as in~\cref{eq:local}, which allows the newly generated points to be dragged, leading to structure deformation. For the rigid dragging, the part $P$ is fixed to encourage an integral movement. Second, we introduce a rigid constraint loss to enforce the rigid moving without the deformation:
\begin{equation}
    L_{\text{rigid}} = \sum_{i \in I(P)} | \| p_s - p_i \|_2 - \| p_s^{*} - p_i^{*} \|_2 |,
\end{equation}
where $p_s^{*}$ and $p_i^{*}$ are the initial position of the source point and its neighboring points, respectively. $L_{\text{rigid}}$ encourages the distance between the neighboring points and the source point to remain unchanged, leading to integral movement.

In general, the user can flexibly combine the deformable and rigid dragging operations by switching the above adjustments and accordingly modifying the hyperparameters (e.g., the movement step $\alpha$ and the loss weight for $L_{\text{rigid}}$).

\subsubsection{Semantic Editing}
\label{sec:semantic}
Sometimes the users may desire to interact with the generation process by simple text prompts. To achieve this, we further propose a semantic editing operation. Specifically, we can first select a part, and then input a new text prompt (e.g., make the wings burning) to compute the SDS loss, and optimize the parts correspondingly to match the newly added semantic features. This operation can be better accomplished with an Interactive SDS loss discussed in the following section.

\subsubsection{Interactive SDS Loss}
\label{sec:loss}
Once the above modifications are finished, we propose to utilize an interactive SDS loss to efficiently optimize the Gaussian blobs. First, observing that the user may use part-level interactions in most cases, we do not need to render the whole view of the object. Instead, we adopt an adaptive camera zoom-in strategy by putting the camera near the modified region and computing the SDS loss using local renderings and user-modified text prompts (\cref{sec:semantic}), resulting in significantly improved optimization efficiency. Furthermore, since we can interact with the entire generation process, we make the denoising step $t$ in the SDS loss adjustable for the user to adapt to different stages. For example, in the early stage with noisier shapes, $t$ can be set to a large number (e.g., $0.98$), while when we aim to fine-tune some parts, $t$ can be set to a small number (e.g., $0.3$) to avoid drastic changes. 

\begin{figure*}[htbp]
  \setlength{\abovecaptionskip}{0.1cm}
  \setlength{\belowcaptionskip}{-0.1cm}
  \centering
    \includegraphics[width=1\linewidth]{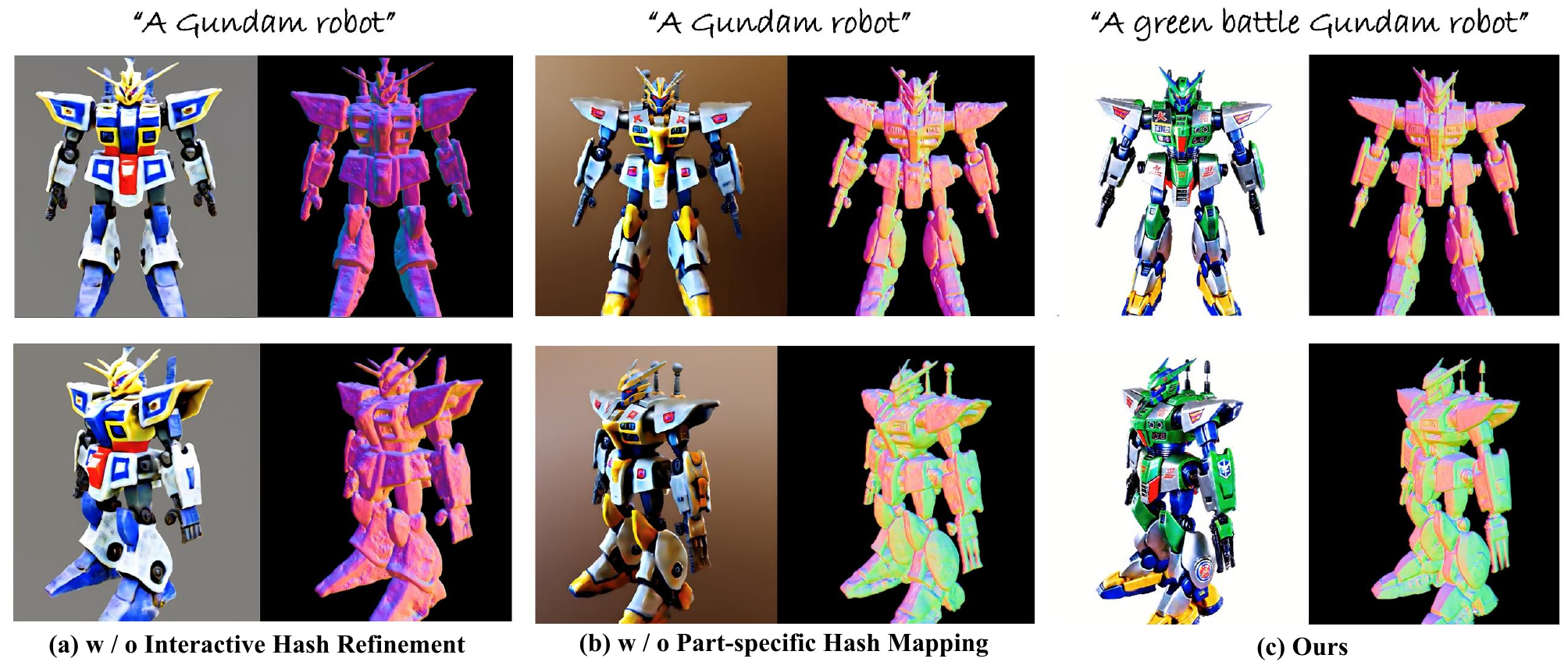}
  \caption{The effectiveness of Interactive Hash Refinement. (a) Results with the original hash table length and resolution from InstantNGP. (b) Results with more refined hash tables and features while without part-specific hash mapping. (c) Results with complete Interactive Hash Refinement module.}
  \label{fig:refine}
\end{figure*}

\subsection{Refinement with InstantNGP}
\label{sec:stage2}
Once obtaining the generated Gaussian blobs $\mathcal{E}$ in Stage-I, we can first convert it to the InstantNGP~\cite{mueller2022instant} representation by a NeRF distillation operation~\cref{sec:dist} and then use the Hash Refinement module~\cref{sec:hash} for further fine-tuning.

\subsubsection{NeRF Distillation}
\label{sec:dist}
Different from the reconstruction task, we find that the Gaussian blobs cannot generate fine-grained 3D objects under the supervision of the SDS loss, often with long trip noises and artifacts. We attribute this to the unstable optimization process for independent Gaussian blobs when supervised by unstable SDS losses. In addition, extracting meshes or other geometries from Gaussian Splatting is still unsolved. Therefore, to further improve the quality and extract geometry, we convert the Gaussian blobs into the InstantNGP representation.
Specifically, we adopt a simple yet effective distillation approach by supervising random renderings from the InstantNGP with the images rendered by Gaussian Splatting:
\begin{equation}
    L_{\text{distill}} = \frac{1}{M}\sum\limits_{c \sim C(\theta, \phi)}\|\mathcal{V}(\mathcal{F},c) - \mathcal{R}(\mathcal{E}, c)\|_1,
\end{equation}
where $c$ is the camera pose derived from a predefined pose distribution $C(\theta, \phi)$; $\mathcal{R}$ is the Gaussian Splatting rendering process; $\mathcal{V}$ and $\mathcal{F}$ are the volumetric rendering and parameters of InstantNGP, respectively.

\subsubsection{Interactive Hash Refinement}
\label{sec:hash}
Upon obtaining the converted InstantNGP representation $\mathcal{I} = \left \{ \mathcal{F}, \mathcal{H} \right \}$, where $\mathcal{H} = \left \{ H_k \right \}_{k=1}^{L}$ contains $L$-level hash tables and $\mathcal{F} = \left \{ F_k \right \}_{k=1}^{L}$ contains corresponding $L$-level features, we aim to interactively refine $\mathcal{I}$ to further improve the generation quality. To achieve this, a simple way is to select the unsatisfying regions and put the camera near these areas for further refinement like what we did in~\cref{sec:semantic}. However, unlike Gaussian Splatting, which can grow new blobs with infinite resolutions and numbers, the informative features stored in InstantNGP have limited capacity (e.g., finite grid resolution and hash table length), leading to refinement bottleneck. Furthermore, InstantNGP does not handle hash conflicts so that different parts may enjoy the same features, resulting in the degeneration of other parts when fine-tuning one part. 

To handle these problems, we propose an Interactive Hash Refinement module, which fixes the original InstantNGP and adds new learnable residual features to the interested regions, overcoming the hash conflict problem by introducing part-specific hash mapping.~The Interactive Hash-Refinement strategy (i) adaptively adds informative features to different regions. For example, we can add hash tables with more levels and features for highly complex regions, while adding fewer residuals in regions that only need to remove artifacts, and the strategy can also (ii) make the model focus on worse regions while freezing the satisfying parts without introducing degeneration.



Specifically, we first extract the binary occupancy grids $O$ with a resolution of $32 \times 32 \times 32$ from the converted $\mathcal{I}$. Given the interested region $Q$ defined by a center $o$ and radius $r$:
\begin{equation}
    Q = \left \{ p | \| p - o \|_2 \leq r, \forall p \in \mathbb{R}^{3} \right \}.
\end{equation}
We intersect $Q$ and $O$ to obtain a part occupancy region $O^{\text{part}} = O \cap Q$. Then we divide $O^{\text{part}}$ to multi-resolution grids and establish a part-specific multi-level hash table set $\mathcal{H}^{\text{part}}$ and learnable features $\mathcal{F}^{\text{part}}$, mapping solely from the local region to local features, avoiding sharing information with other object parts:
\begin{equation}
    f_k = \begin{cases}
    H_{k}^{\text{part}}(p, F_{k}^{\text{part}}), & p \in O^{\text{part}}\\
    0, & p \notin O^{\text{part}}
    \end{cases},
\end{equation}
where $f_k$ denotes the mapped features at level $k$ at position $p$.
Subsequently, we introduce new lightweight MLPs to convert newly added features to residual densities and colors to influence the final renderings. 
After that, we can use the interactive SDS loss to optimize the local regions as discussed in~\cref{sec:loss}.
.




\section{Experiment}
\begin{figure}[htbp]
  \setlength{\abovecaptionskip}{0cm}
  \setlength{\belowcaptionskip}{-0.6cm}
  \centering
  \includegraphics[width=0.9\linewidth]{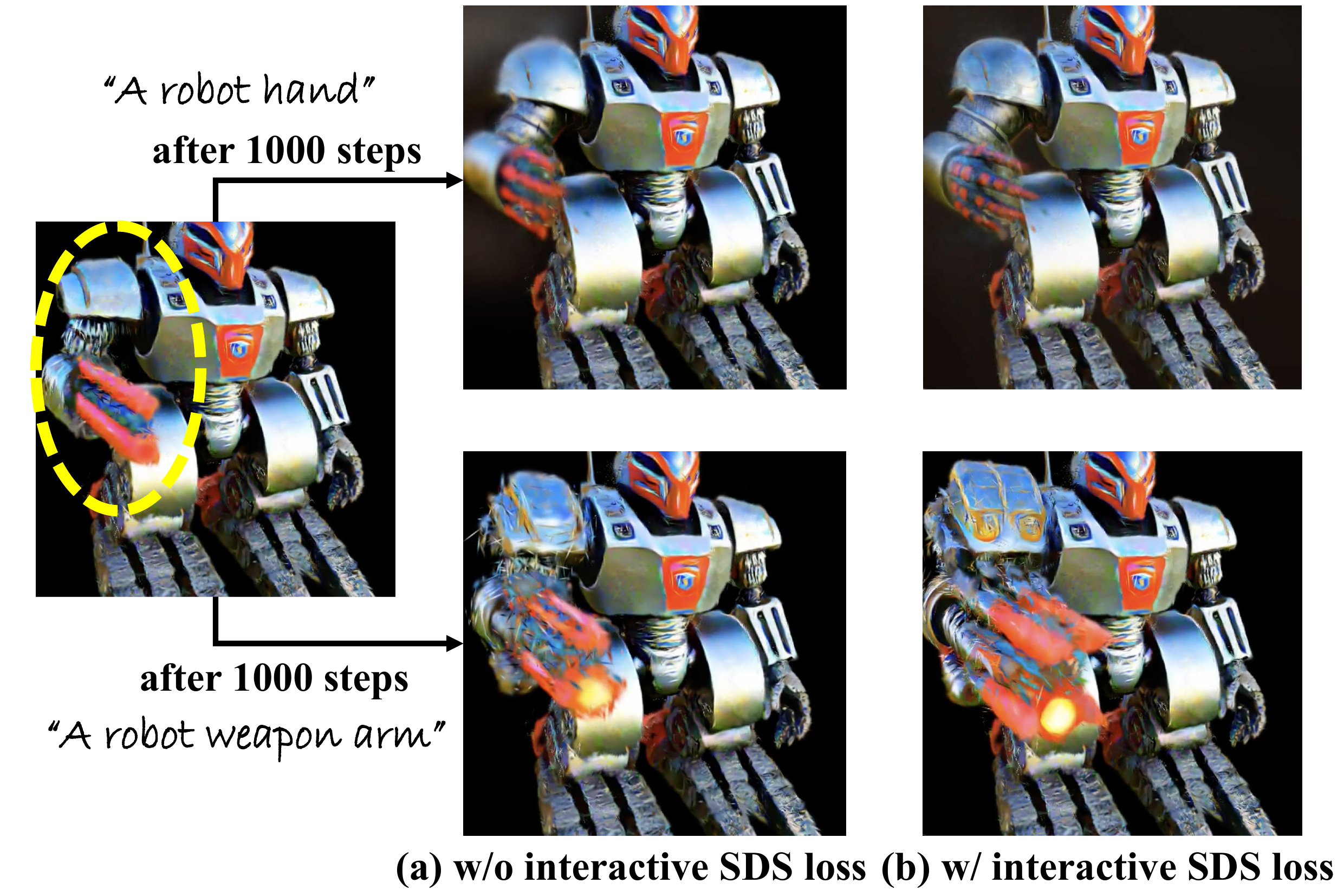}
  \caption{Ablation study of the proposed Interactive SDS loss.}
  \label{fig:loss_ab}
\end{figure}


\subsection{Implementation Details}
\label{sec:inp}
We build \textit{Interactive3D} in two stages with different 3D representations.~We follow~\cite{chen2023text} to implement the Gaussian Splatting-based optimization in Stage I and follow~\cite{mueller2022instant} to implement InstantNGP-based refinement in Stage II. We implement the Hash Refinement module in parallel with CUDA to improve efficiency. The general training step for one object is 20k (10k for Stage I and 10k for Stage II). We conduct our experiments on NVIDIA A100 GPUs. More details can be found in the supplementary materials.

\begin{figure*}[t]
  \setlength{\abovecaptionskip}{0cm}
  \setlength{\belowcaptionskip}{-0.3cm}
  \centering
  \includegraphics[width=0.9\linewidth]{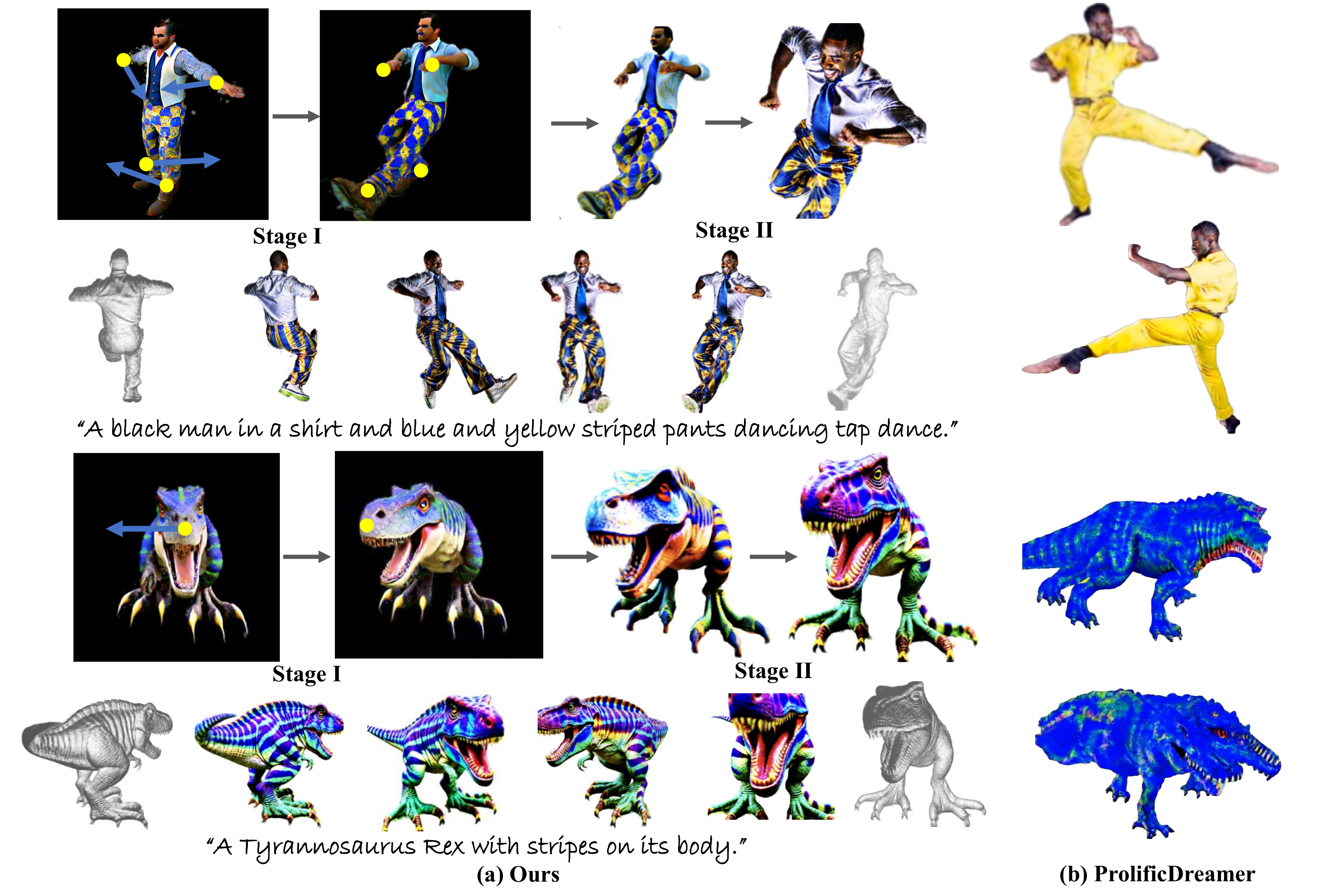}
  \caption{Qualitative results of the rigid dragging operation, demonstrating the effectiveness and controllability of user interactions.}
  \label{fig:results}
\end{figure*}

\subsection{Qualitative Results}
\label{sec:result1}
We show some 3D generation results from $\textit{Interactive3D}$ as well as the interaction process in~\cref{fig:results},~\cref{fig:watermelon}, and~\cref{fig:refine} (c). As shown in the first row of~\cref{fig:results}, we can interrupt the generation process and change the pose of the generated human, and then continue to optimize. It can be observed that the standing 3D man is smoothly converted to a kick-dancing man by directly modifying the Gaussian blobs. The dancing man is then optimized by a few steps to fix some artifacts and holes after the interaction, and it is converted to InstantNGP for further refinement. Compared with the results directly generated from current state-of-the-art methods, our generated objects are more controllable with better geometry and texture. For instance, 
ProlificDreamer has totally wrong geometry and mismatches the text prompt.
In the second row of~\cref{fig:results}, we rigidly drag the head of Tyrannosaurus Rex from looking forward to the right direction, achieving a DragGAN~\cite{tewari2023drag}-style interaction while in 3D. 
In~\cref{fig:refine} (c), we show that by applying our Interactive Hash Refinement module to a coarse 3D object, \textit{a Gundam robot} in this case, we can achieve significant textural and geometric improvements. It is noteworthy that the interactions can be combined in one generation process as shown in~\cref{fig:teaser}. More results can be found in the supplementary material.

\subsection{Quantitative Results}
\label{sec:result2}
Following~\cite{jun2023shap}, we use the CLIP R-Precision to quantitatively evaluate our generated results in~\cref{tab:quantitive}. We use 50 prompts derived from Cap3D~\cite{luo2023scalable} and compare the results with other methods. \textit{Interactive3D} achieves the highest CLIP R-Precision by interactive generation, which demonstrates the strong controllability of our method. Meanwhile, we can achieve highly efficient 3D generation, because of the following two reasons: (i) we adopt the fast Gaussian Splatting in Stage I and utilize its results to initialize the optimization in Stage II which speeds up the convergence; (ii) the interactions are incorporated into some checkpoint of the original optimization process without the need for extra training steps, we achieve efficient 3D generation.

\begin{table}[t]
\setlength{\abovecaptionskip}{0cm}
\setlength{\belowcaptionskip}{0cm}
  \centering
  \caption{Quantitative comparison on the CLIP R-Precision~\cite{jun2023shap}.}
  \begin{tabular}{cccc}
    \toprule
    Method & R-Precision & Average Time \\
    \midrule
    DreamFusion~\cite{poole2022dreamfusion} & 0.67 & \textbf{1.1h} \\
    ProlificDreamer~\cite{wang2023prolificdreamer} & 0.83 & 3.4h \\
    Ours        & \textbf{0.94} & \textbf{50min}\\
    \bottomrule
  \end{tabular}
  \vspace{-.4em}
  \label{tab:quantitive}
\end{table}

\begin{figure}[htbp]
  \setlength{\abovecaptionskip}{0cm}
  \setlength{\belowcaptionskip}{-0.6cm}
  \centering
  \includegraphics[width=1.0\linewidth]{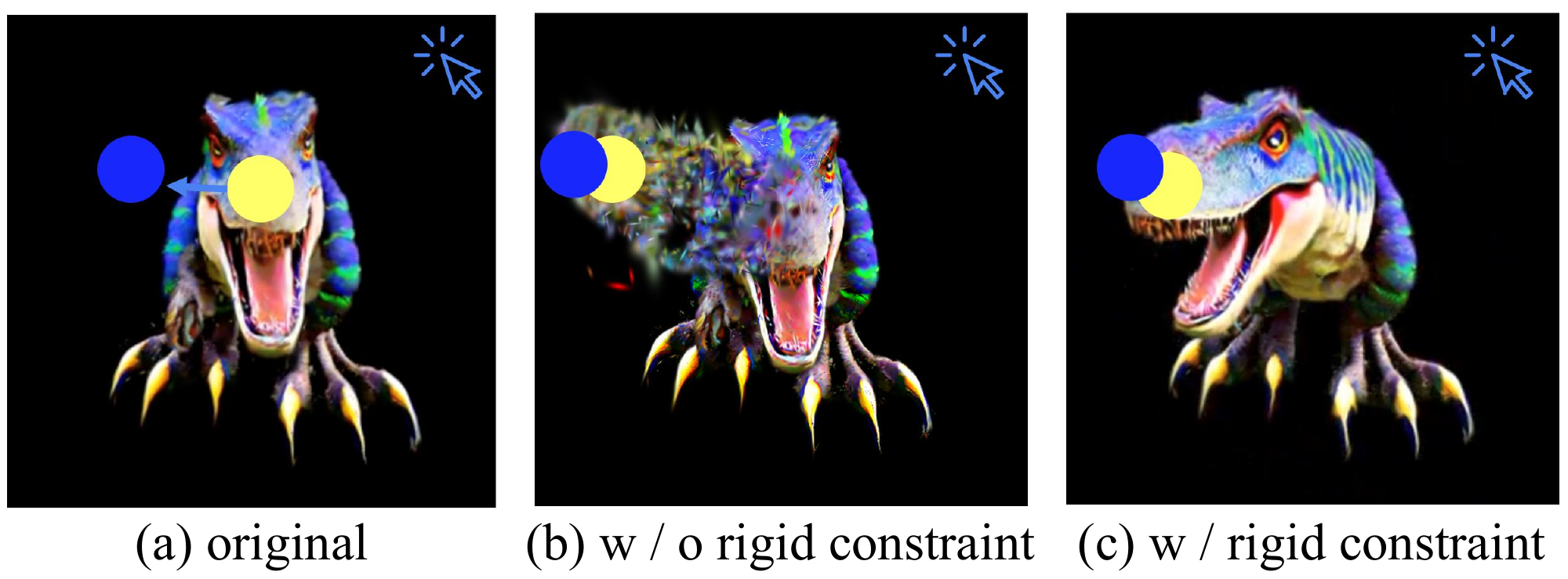}
  \caption{Ablation study of the proposed Rigid Constraint loss.}
  \label{fig:rigid_ab}
\end{figure}

\subsection{Ablation Studies}
\label{sec:ab}
Now, we investigate the impact of the Rigid Constraint loss, the Interactive SDS loss, and the Hash Refinement Module.
More ablations can be found in the supplementary material.

\vspace{-.7em}
\paragraph{Effect of Rigid Constraint Loss}
As shown in ~\cref{fig:rigid_ab}, the Rigid Constraint loss plays a vital role in ensuring the local structure unchanged during the dragging. With this constraint, the local structure can be well maintained during the dragging while eliminating this loss leads to deformation.

\vspace{-1em}
\paragraph{Effect of Interactive SDS Loss}
We ablate the Interactive SDS loss in~\cref{fig:loss_ab}. 
Given a robot in the representation of Gaussian blobs, we aim to change its original arm to other formats. The results show that we achieve better-modified results by using the Interactive SDS loss after the same number of training steps compared with the baseline. 
\vspace{-1em}
\paragraph{Effect of Interactive Hash Refinement Module}
We ablate the Interactive Hash Refinement Module and the effectiveness of part-specific hash mapping in~\cref{fig:refine}. Refer to~\cref{fig:refine} (a), it shows that the generation quality is blurry and lacks details with only the original SDS optimization, and the details and artifacts cannot be improved even with extremely long optimization steps (e.g., 100k training iterations). The integration of the proposed interactive refinement enhances the realism of models by correcting unreasonable components, such as the robot head in~\cref{fig:refine} (b). The proposed part-specific hash mapping further advances the texture and geometric precision of the model dramatically, as shown in \cref{fig:refine} (c). Furthermore, the part-specific hash mapping is also important to avoid region conflicts, leading to more fine-grained generation results.
\vspace{-.7em}
\section{Conclusion}
\vspace{-.7em}
In this paper, we proposed \textit{Interactive3D}, a novel framework for controllable and high-quality 3D generation.
Interactive3D works in two main stages, using different 3D representations. In Stage I, we use Gaussian Splatting representation which allows users to flexibly change the model by adding or removing parts, transforming shapes, and making semantic editing. In Stage II, we first convert the Gaussian blobs to InstantNGP by fast NeRF distillation, and then propose a novel Interactive Hash Refinement to further improve the quality and extract 3D geometry. 

\noindent\textbf{Acknowledgments:}
This work was supported in part by Research Grants Council (RGC) of the Hong Kong SAR under grant No.~26202321, HKUST startup fund No.~R9253, and CUHK Direct Grants (RCFUS) No.~4055189.~We also gratefully acknowledge the support of SenseTime.


{
    \small
    \bibliographystyle{ieeenat_fullname}
    \bibliography{main}
}
\clearpage
\setcounter{page}{1}
\maketitlesupplementary

\begin{strip}\centering
\includegraphics[width=\textwidth]{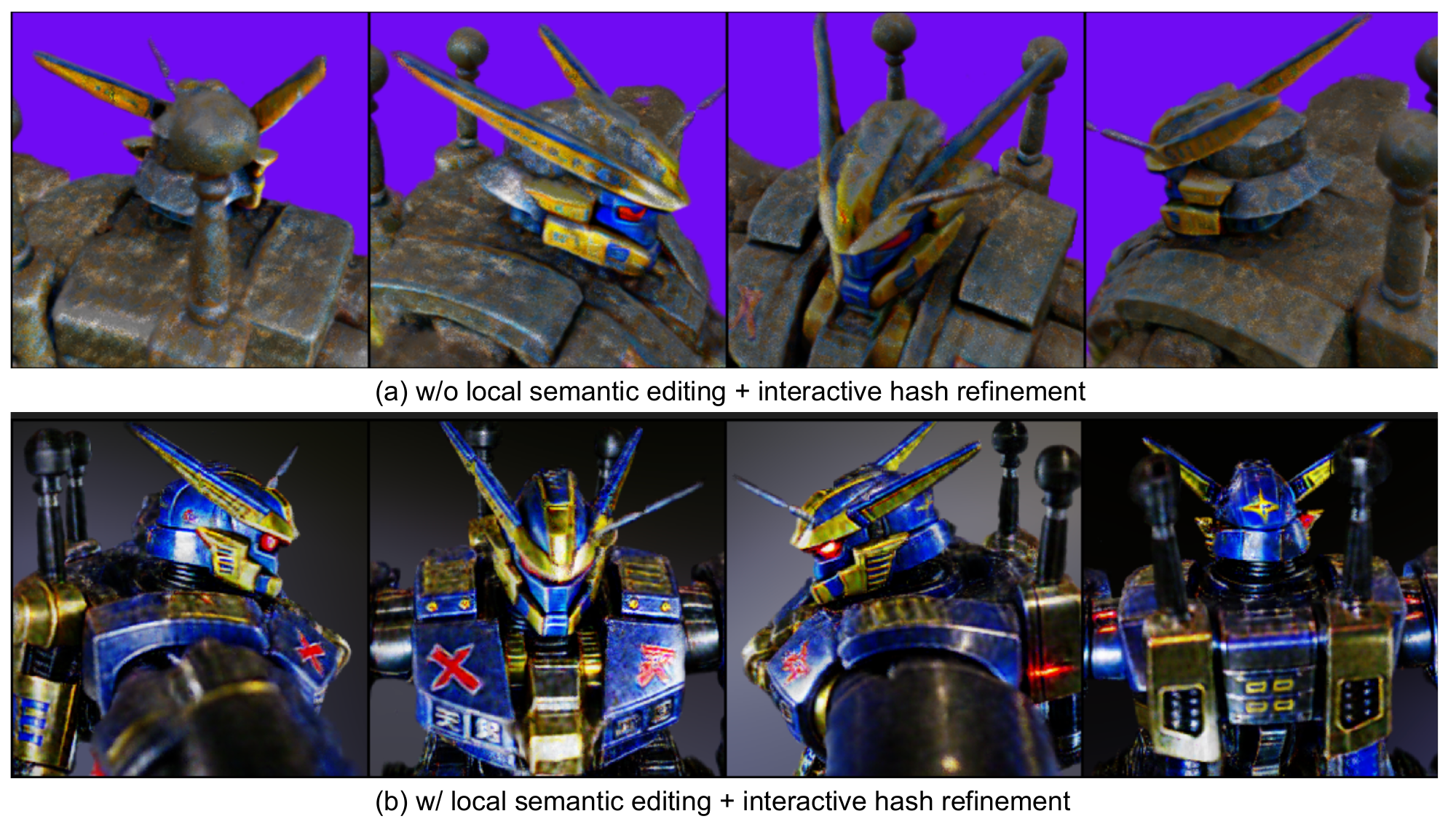}
\captionof{figure}{\textbf{The effectiveness of the combination refinement.} Refine \textit{a Gundam robot}.
\label{fig:more_ab}}
\end{strip}


\begin{strip}\centering
  \includegraphics[width=1.0\linewidth]{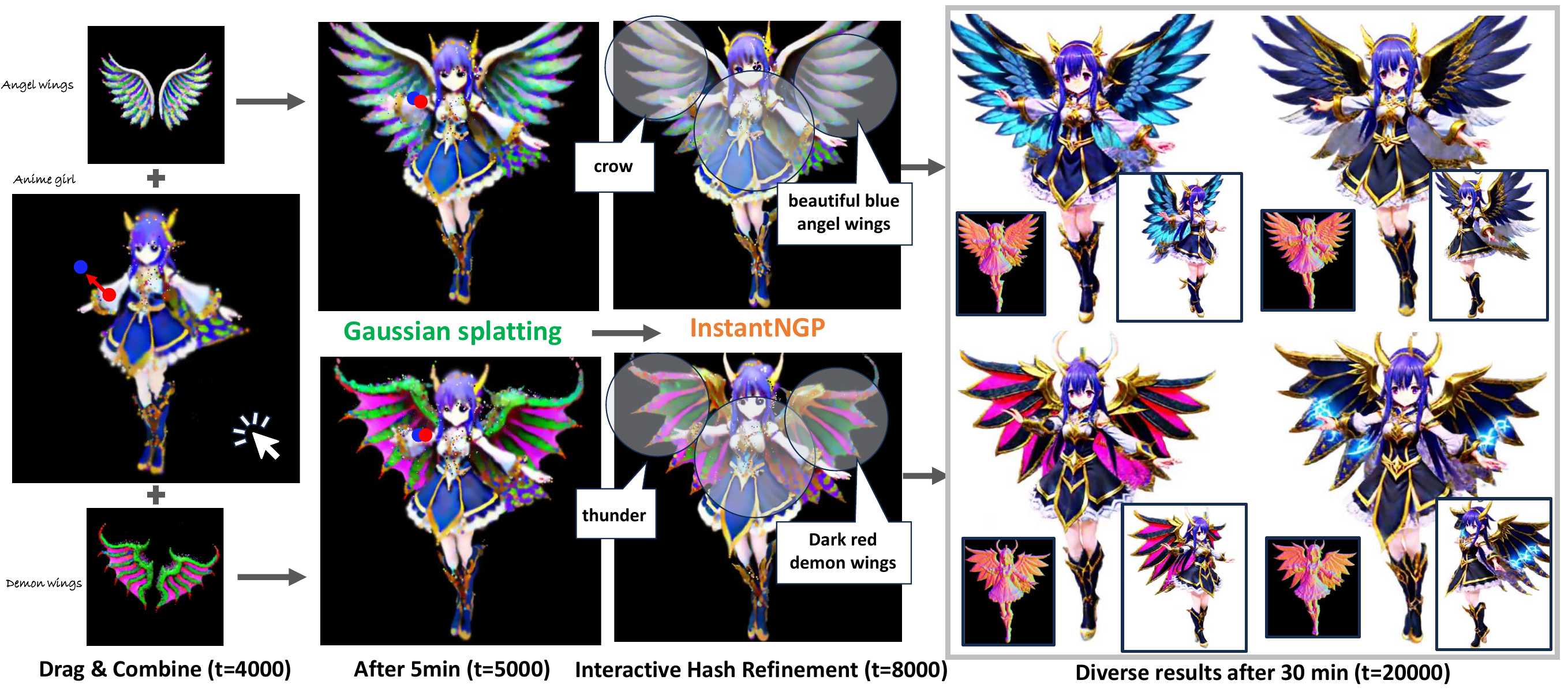}
  \captionof{figure}{\textbf{Qualitative generation results of the proposed Interactive3D.} We show the interaction process of generating angel girls.}
  \label{fig:girl_edit}
\end{strip}

\begin{strip}\centering
  \setlength{\abovecaptionskip}{0.1cm}
  \setlength{\belowcaptionskip}{-0.4cm}
  \includegraphics[width=0.9\linewidth]{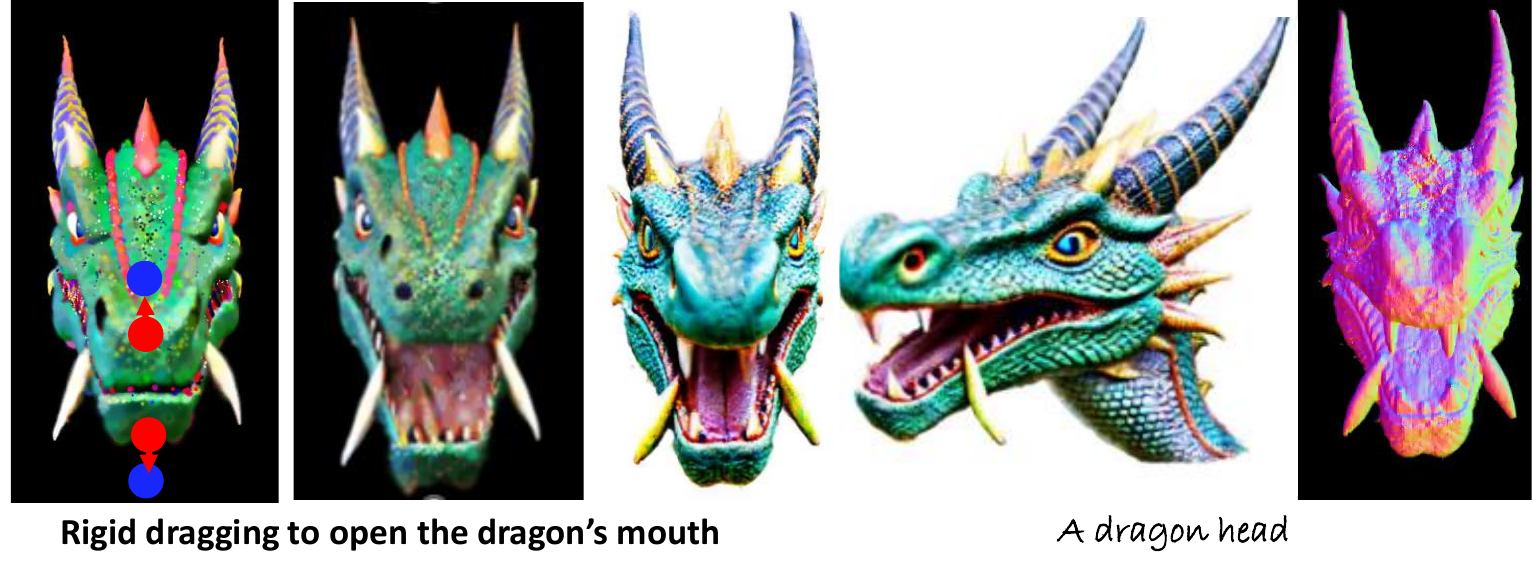}
  \captionof{figure}{\textbf{Generation results of the rigid dragging operation}. We use the rigid dragging to open the dragon's mouth.}
  \label{fig:dragon}
\end{strip}

\begin{strip}\centering
  \setlength{\abovecaptionskip}{0.1cm}
  \includegraphics[width=0.9\linewidth]{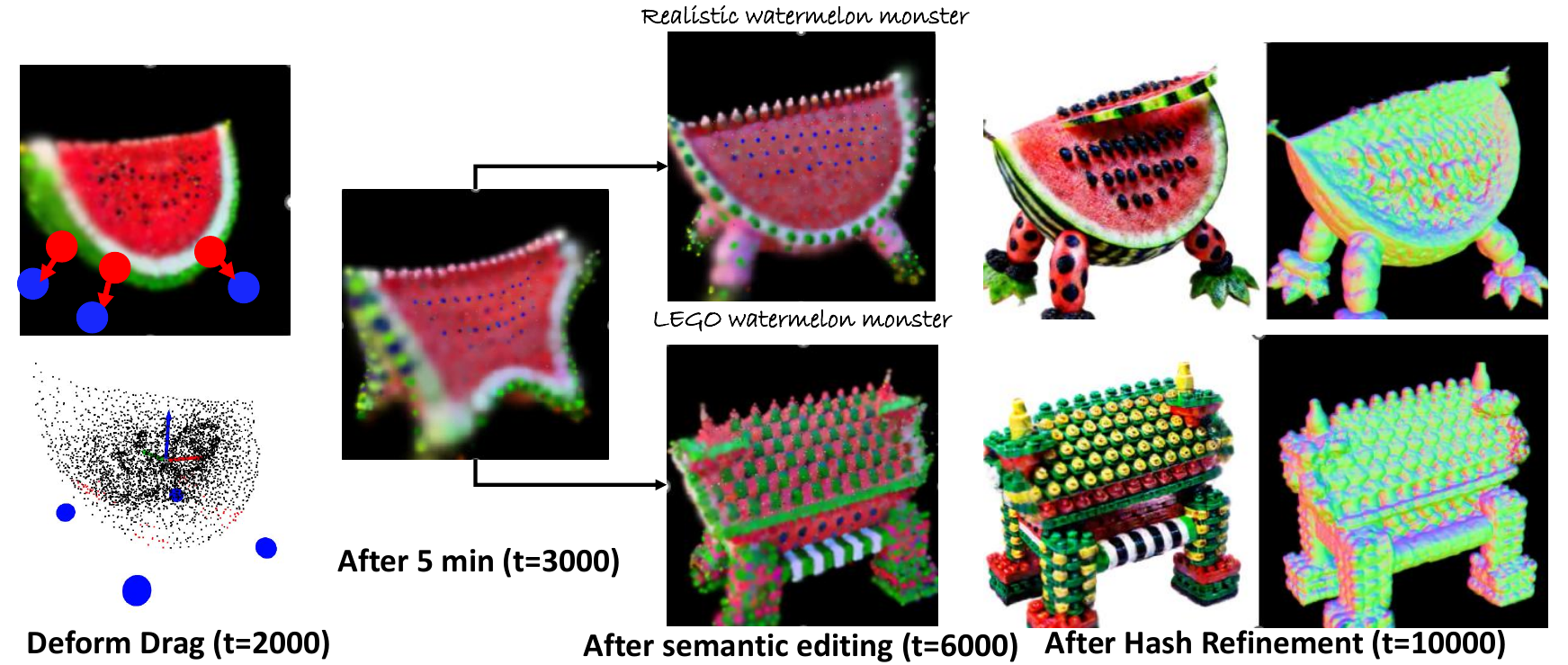}
  \captionof{figure}{\textbf{Results of the proposed deformable dragging and semantic editing operations.} Our Interactive3d allows for flexible geometric deformations and semantic modifications of 3D objects.}
  \label{fig:watermelon}
\end{strip}

In the supplementary material, we first introduce more implementation details in~\S~\ref{sec:more_inp}, and then provide more results in~\S ~\ref{sec:more_results} and the online \href{https://interactive-3d.github.io/}{project page}. We also provide additional ablation studies in~\S ~\ref{sec:more_ab}.

\subsection{Implementation Details}
\label{sec:more_inp}
We train our framework on one NVIDIA A100 GPU for 20k steps. We use the AdamW optimizer with $\beta = (0.9, 0.999)$ and the weight decay $=0.01$. For Stage I, we utilize $4096$ points sampled from Shap-E~\cite{jun2023shap} to initialize the Gaussian blobs following~\cite{chen2023text}. The learning rate for position, scale, rotation, color, and alpha are set to $0.005, 0.003, 0.003, 0.01, 0.003$, respectively. We use Stable Diffusion 2.1~\cite{metzer2022latent} as our 2D prior. For Stage II, by default, we set the number of levels of the refinement hash table to $8$, and set the feature dimension of each position to $2$. The capacity of each refinement hash table is set to $2^{19}$. We render $256 \times 256$ images during training and also use Stable Diffusion 2.1 to provide guidance. Notably, we follow~\cite{poole2022dreamfusion} to use the orientation loss as a regularization.

\begin{table}[t]
\setlength{\abovecaptionskip}{0cm}
\setlength{\belowcaptionskip}{-0.2cm}
  \centering
  \caption{Efficiency of using the proposed Interactive SDS loss.}
  \begin{tabular}{cc}
    \toprule
    Interactive SDS loss & Average Time \\
    \midrule
    w/o & 1.8h \\
    w & \textbf{50min} \\
    \bottomrule
  \end{tabular}
  \vspace{-1em}
  \label{tab:sds_loss}
\end{table}

\begin{figure*}[!t]
  \centering
  \includegraphics[width=0.8\linewidth]{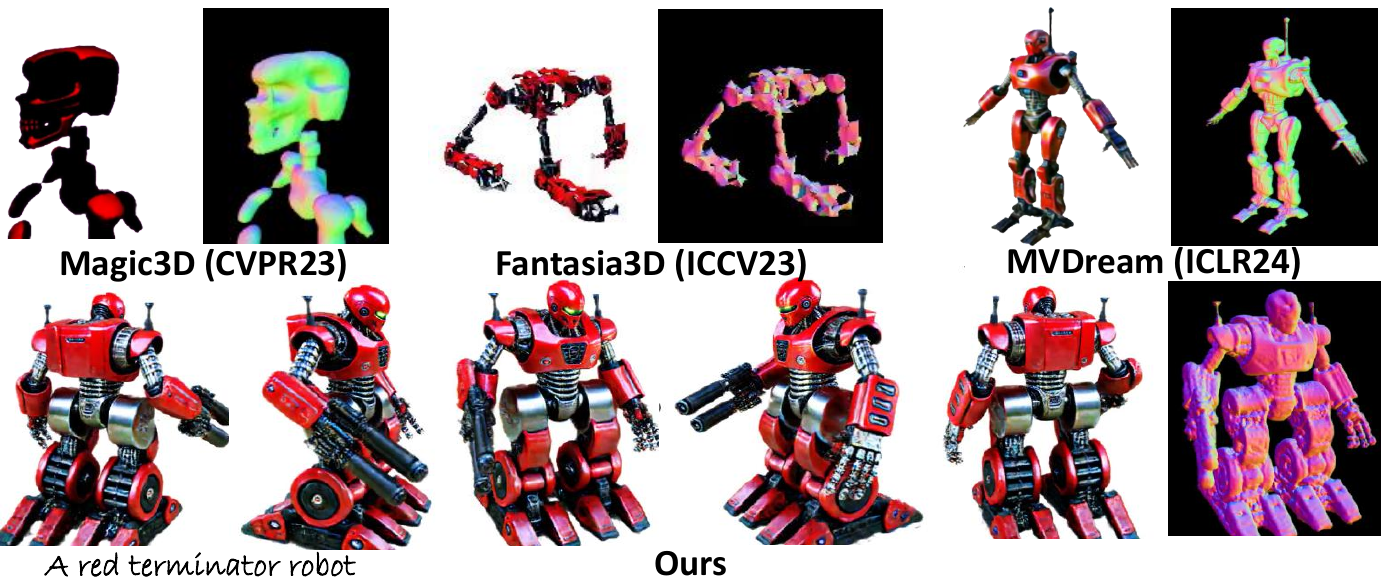}
  \caption{\textbf{More comparisons}. We compare Interactive3D with previous state-of-the-art text-to-3D methods.}
  \label{fig:more_comp}
\end{figure*}

\begin{figure*}[!t]
  \setlength{\abovecaptionskip}{0cm}
  \setlength{\belowcaptionskip}{-0.4cm}
  \centering
  \includegraphics[width=0.8\linewidth]{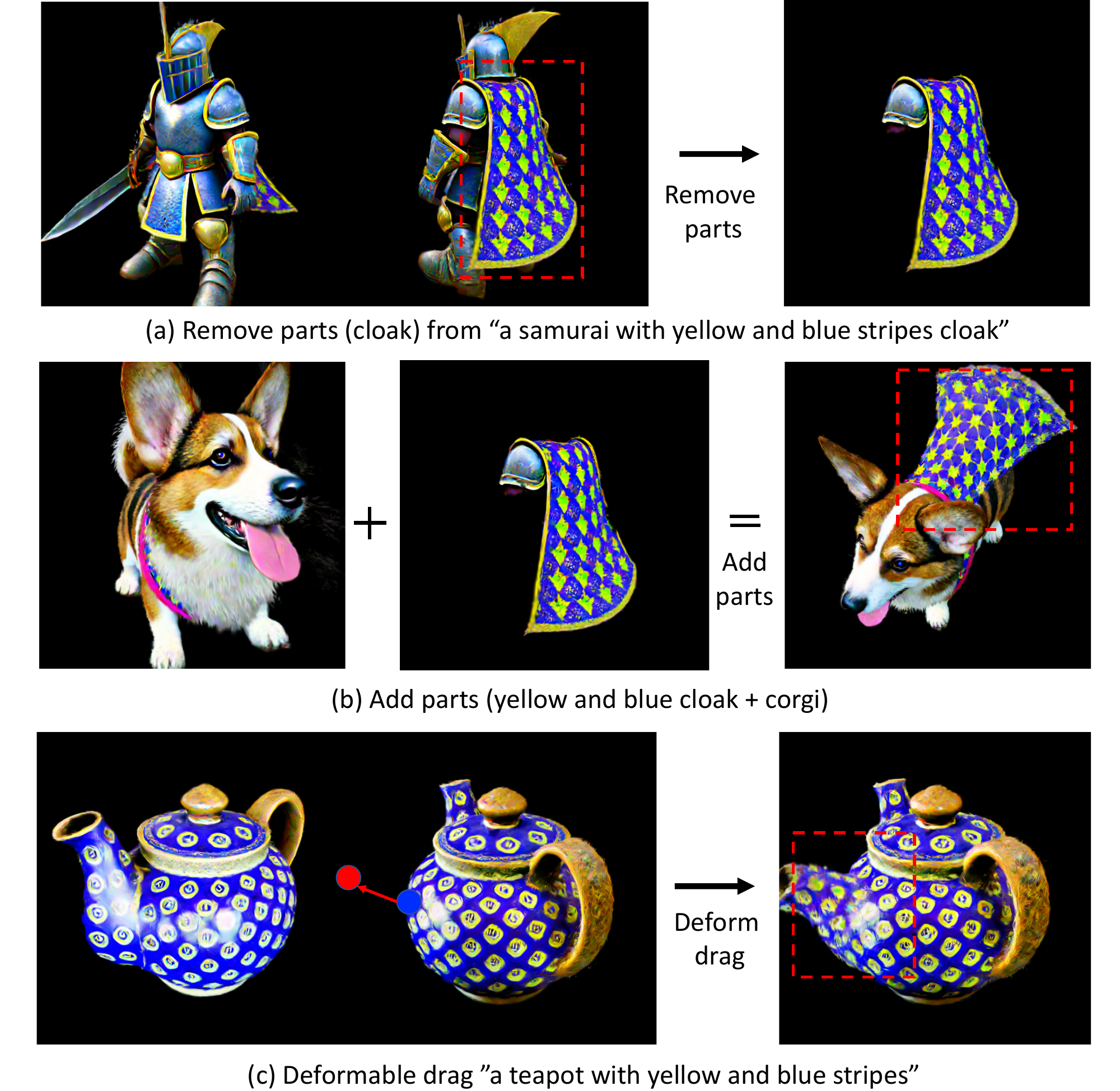}
  \caption{\textbf{Results of Interactive3D Stage I}. We show more results of dragging, removing, and adding parts. Note that all the above results after interaction (right side) are from Stage I without Stage II's refinement.}
  \label{fig:more_drag}
\end{figure*}

\subsection{More Results}
\label{sec:more_results}
In this section, we give more visualization results of our framework including dragging, removing and adding parts, and interactive hash refinement. 
In~\cref{fig:girl_edit}, we show the final results from our entire generation framework (Stage I and Stage II) and the interaction process for generating different angel girls. It can be observed that the generated two anime girls have high-quality texture and accurate geometry. In~\cref{fig:dragon}, we can open a dragon's mouth with the rigid dragging operation. 
In~\cref{fig:watermelon}, we use the deformable dragging operation to create a watermelon monster and utilize the semantic editing operation to customize the style. 
In ~\cref{fig:more_drag}, we show more generated results of Stage I after user's interaction such as dragging, removing, and adding parts. \cref{fig:more_drag} (a) shows that we can accurately segment (remove) the cloak from the samurai. Furthermore, the segmented cloak is combined with a corgi in~\cref{fig:more_drag} (b). In~\cref{fig:more_drag} (c), we create a new spout from the teapot by deformable dragging. \cref{fig:more_comp} shows more comparisons.

\setlength{\columnsep}{5pt}%
\begin{wraptable}{r}{4.4cm}
  \setlength\tabcolsep{1pt}
  \renewcommand\arraystretch{0.8}
  \setlength{\abovecaptionskip}{-0.0mm}  
  \setlength{\belowcaptionskip}{8pt}
  \centering
  \vspace{-8mm}
  \caption{User studies (22 pepole).}
  \label{tab:user_study}
  \footnotesize
    \begin{tabular}{c c c}
    \toprule
    Method & Preference$\uparrow$ & Avg attempts$\downarrow$ \\
    text-only & 4.5\% & 2.3 \\
    \midrule
    Ours &  \textbf{95.5}\% & \textbf{1.4} \\
    \bottomrule
    \end{tabular}
    \vspace{-4mm}
\end{wraptable}%

In addition, we provide user studies in \cref{tab:user_study}. Most users agree that interactive3D is convenient and needs fewer attempts. They can create objects better match their expectations than solely using text prompts as in previous methods.

\subsection{More Ablation Results}
\label{sec:more_ab}
\paragraph{Effect of Combination Refinement} As mentioned in the main paper, semantic editing and interactive refinement can be combined to improve the quality. As demonstrated in~\cref{fig:more_ab}, users can perform semantic editing and refinement on any local detail by the combination refinement.
\newpage
\paragraph{Effect of Interactive SDS Loss}
We provided qualitative ablation experiments on the Interactive SDS Loss in~\cref{fig:loss_ab} from the main paper. We provide a quantitative comparison for the Interactive SDS loss here. As shown in~\cref{tab:sds_loss}, the Interactive SDS loss significantly improves the efficiency of our whole pipeline.

\subsection{Limitations}
\label{sec:limitations}
One limitation of our Interactive3D method is its susceptibility to failure under conditions of excessive and unreasonable manipulation (e.g., in the case of unreasonable dragging operations). This vulnerability highlights the method's reliance on user discretion for optimal outcomes. Additionally, Interactive3D is built upon the foundations of current generative techniques, inheriting their common challenges, including issues related to color saturation. This dependence means that Interactive3D is not immune to the intrinsic limitations of the underlying technologies it utilizes.

\end{document}